\documentclass{iise}

\usepackage{bigstrut, booktabs, multirow}
\usepackage[nolist]{acronym}
\usepackage[belowskip=-5pt,aboveskip=0pt]{caption}

\usepackage{array,tabularx,calc} 

\usepackage{siunitx}

\conference{Proceedings of the IISE Annual Conference \& Expo 2024\\
A. Brown Greer, C. Contardo, J.-M. Frayret, eds.}

\title{\titlesize Evaluating the Impact of Vaccine Hesitancy on the Allocation of Vital Resources During COVID-19 Pandemic}

\author{
Hieu Bui$^1$, Sandra D. Eksioglu$^1$, Ruben A. Proano$^2$ \\
\vspace{0.2cm}
$^1$Department of Industrial Engineering, \\
University of Arkansas, Fayetteville, AR 72701  \\ \vspace{0.2cm}
$^2$Department of Industrial and Systems Engineering,\\
Rochester Institute of Technology, Rochester, NY 14623
}

\authorlist{Bui, Eksioglu, Proano}
\abstractID{7318}

\begin{document}

\maketitle

\begin{abstract} 

{\small The COVID-19 pandemic highlighted significant challenges in the allocation of vital healthcare resources. Existing epidemiological models, specifically compartmental models, aimed to predict the spread of the COVID-19 virus and its impact on the population, but they overlooked the influence of \ac{VH} on disease dynamics, including the expected number of hospitalizations and fatalities. We propose improvements to the \ac{SEIR} model for COVID-19 by incorporating the influence of vaccination, \ac{VH}, and resource availability on the disease dynamics. We collect publicly available data and perform data analysis to capture \ac{VH} dynamic changes over time and develop scenario paths for \ac{VH}. We simulate the proposed compartmental model for each \ac{VH} path to explain the impacts of public attitudes toward vaccination, the impacts of healthcare resources on patient outcomes, and the timing of vaccination rollout on the progression and severity of the epidemic. Our analysis demonstrates that reducing \ac{VH} improves health outcomes, reinforcing the importance of addressing \ac{VH} to curb the spread of infectious diseases. Our results show that adequate levels of critical healthcare resources are crucial for minimizing fatalities and also highlight the life-saving impact of timely and effective vaccination programs.}
\end{abstract}

\vspace{-0.2in}

\section*{Keywords}
compartmental model, COVID-19, data analytics, healthcare resources, vaccine hesitancy 

\vspace{-0.1in}

\section{Introduction}
\label{sec:intro}

The emergence of infectious diseases with high transmission rates, such as COVID-19, emphasizes the critical role of strategic public health interventions to limit the spread of the disease, and reduce hospitalizations and fatalities, such as vaccination campaigns, social distancing, and the strategic allocation of vital resources \cite{moghadas_2021}. The success of vaccination campaigns is compromised by \acf{VH}, which is the resistance or refusal to be immunized despite the availability of vaccines. \ac{VH} is a  barrier to achieving widespread immunity, challenging public compliance, and complicating healthcare resource allocation\cite{blasioli_2023}.

Strategic public policies are essential to manage epidemics. The effectiveness of these policies depends on the accurate prediction of the spread of the disease. Compartmental epidemiological models, such as \ac{SIR} and \acf{SEIR} models, have been instrumental in this regard. The \ac{SIR} model was introduced in the early 20th century. It divides the population into \textbf{S}usceptible, \textbf{I}nfectious, and \textbf{R}ecovered compartments to mimic the changes in the health state of a homogenous population affected by an infectious disease \cite{kermack_1927}. The \ac{SEIR} model extends the \ac{SIR} by including an \textbf{E}xposed state to account for individuals exposed to a virus that show symptoms and become infectious after an incubation period. Reviews by \cite{kong_2022, xiang_2021} show that compartmental models have been widely used to study COVID-19, providing a good representation of its transmission dynamics. Traditional \ac{SEIR} models are deterministic (i.e., assume constant transition rates between compartments); however, several model adaptations are stochastic and capture the inherent randomness in disease transmission. For example, Kucgarski et al. \cite{kucharski_2020} utilized stochastic transmission dynamic models to estimate the transmission rate of COVID-19. Kretzschmar et al. \cite{kretzschmar_2020} applied stochastic mathematical models to estimate effective reproduction numbers, accounting for delays in testing and isolation. Despite the variety of available compartmental models, there is a scarcity of models that explicitly analyze the impact of stochastic \ac{VH} on the spread of the disease and the demand for healthcare resources. 

The compartmental epidemiological models are solved via systems of \acp{ODE} to track population dynamics between different compartments over time. For complex models such as those for COVID-19, closed-form solutions of these \acp{ODE} are impractical, necessitating numerical methods within simulation frameworks. These simulations numerically approximate \ac{ODE} solutions and can help clarify the nuances of disease propagation and the potential impact of different interventions for short time intervals. However, current simulation frameworks are often limited in scope, focusing on specific scenarios that do not capture the entire spectrum of stochastic parameters. This limitation can lead to a narrow understanding of potential outcomes. Hence, there is a need for comprehensive models that explore a wider range of variables and uncertainties.

Our study seeks to fill these gaps by presenting a compartmental model designed specifically for COVID-19, which incorporates the stochasticity of \ac{VH} and constraints on healthcare resource access. We analyze the \ac{VH} data provided by surveys to create a comprehensive spectrum of \ac{VH} trajectories (i.e., scenario paths). We use the model to answer these research questions relevant to the COVID-19 pandemic: {\bf RQ1:} \emph{How do changes in public attitudes towards vaccination affect epidemic outcomes?} {\bf RQ2:} \emph{How do healthcare resources impact patient outcomes during a pandemic?} {\bf RQ3:} \emph{How do vaccine rollout times affect the progression and severity of the epidemic?}

Finally, this research sets the foundation for more sophisticated mathematical models, such as \ac{MSP}, aiming to cultivate a healthcare strategy that is flexible and responsive to the spread of the disease and people's attitudes toward vaccination.

\section{Methodology}
\label{sec:method}

The \ac{SEIR} model divides the population into four compartments: Susceptible ($\pmb{S}$), Exposed ($\pmb{E}$), Infectious ($\pmb{I}$), and Recovered ($\pmb{R}$). Individuals transition through these compartments over time. The transition rate between compartments depends on the specific characteristics of the infectious disease. Section~\ref{sec:sveihr} focuses on our proposed compartmental model that extends the \ac{SEIR} model for COVID-19 to capture the impact of vaccine availability, \ac{VH} stochasticity, and availability of healthcare resources on the spread of the disease and health outcomes. Our numerical analysis indicates that \ac{VH} in the U.S. changed over time due to different public health interventions.  Section~\ref{sec:uncertainty-paths} summarizes our approach to model uncertainty of \ac{VH}.

\subsection{COVID-19-Specific Compartmental Model}
\label{sec:sveihr}

\begin{figure}[htb]
    \centering
    \includegraphics[width=\textwidth]{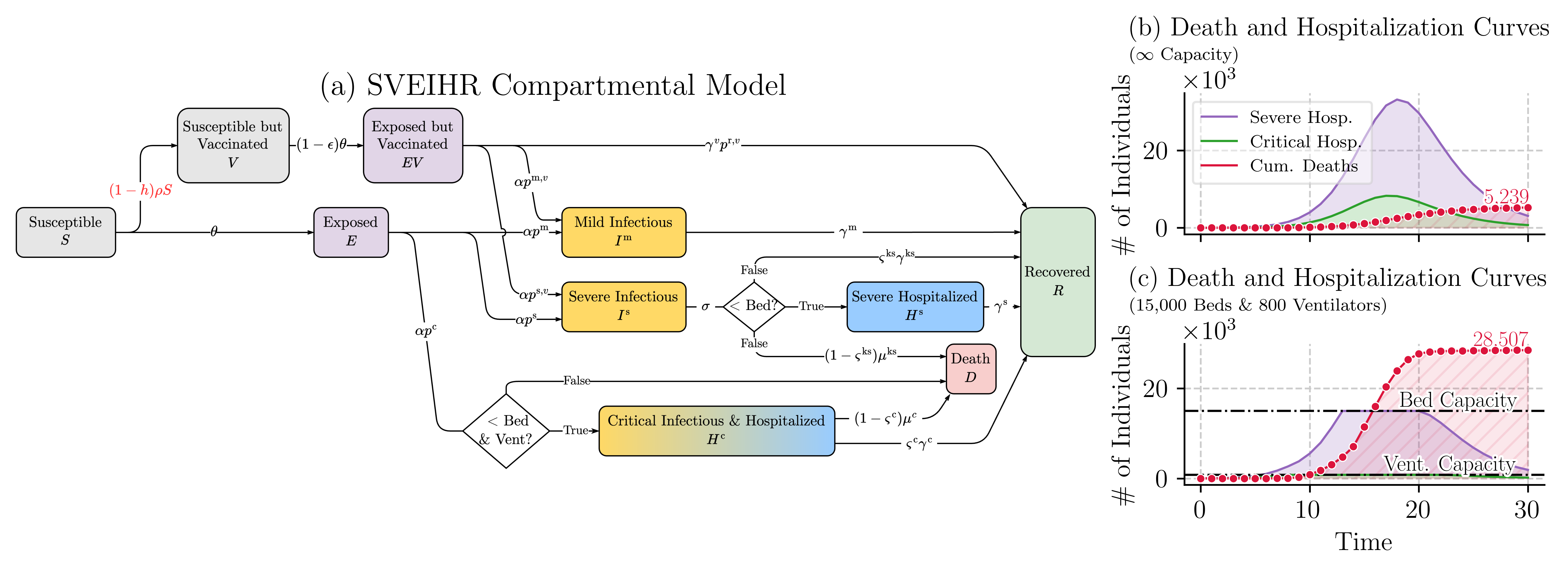}
    \caption{(a) \ac{SVEIHR}: the proposed COVID-19 compartmental model. (b) and (c) illustrate the impact of healthcare capacity on the number of deaths and hospitalizations for a population of size 1.2M.}\label{fig:sveihr}
\end{figure}

The proposed model called \ac{SVEIHR}, summarizes the dynamic progression of COVID-19 in the presence of vaccines (see Figure ~\ref{fig:sveihr}(a). Thus, unlike in the \ac{SEIR} model, individuals in $\pmb{S}$ can transition to compartment $\pmb{E}$ or opt to vaccinate, moving forward to the vaccinated compartment ($\pmb{V}$). This transition is influenced by the expected vaccination  ($\rho$) and \ac{VH} ($h$) rates. We consider $h$ to be a stochastic parameter because \ac{VH} differs by population group based on political, economic, and social standing. \ac{VH} changes over time due to many reasons, including public health campaigns, interventions, and social media commentaries. Since vaccine efficacy ($\epsilon$) is less than 100\%, vaccinated individuals could become exposed. These individuals transition to the compartment ($\pmb{EV}$).

The model simulates different health trajectories post-exposure, based on incubation rates $\alpha$ and severity of the disease. Individuals progress to mild ($\pmb{I^{\text{m}}}$), severe ($\pmb{I^{\text{s}}}$), or critical condition ($\pmb{H^{\text{c}}}$) compartments with probabilities $p^{\text{m}}, p^{\text{s}},$ and $p^{\text{c}}$, respectively. The vaccinated population is less likely to experience severe/critical health outcomes and, if infected, is more likely to recover and move directly to the compartment ($\pmb{R}$). 

The model factors the availability of healthcare resources, such as the number of hospital beds (\textbf{Bed}) and ventilators (\textbf{Vent}), which are crucial in determining patients' health outcomes. To illustrate the significance of bed and ventilator capacity on hospitalizations and fatalities, we conducted two simulations using the \ac{SVEIHR} model within a population of 1.2 million. The first simulation operates under the assumption of unlimited bed and ventilator availability, while the second imposes limitations on these critical resources. The parameters used in these simulations, and other experiments presented in this study, are summarized in Table~\ref{tab:parameters}. Several of the parameters used were derived from the literature \cite{wu_2020, zhou_2020, linton_2020, lim2021case}. We estimated other parameters using a similar procedure as in \cite{carcione2020simulation}. 
The cumulative number of hospitalizations and deaths for the first simulation are presented in {Figure ~\ref{fig:sveihr}(b), and for the second simulation is presented in {Figure ~\ref{fig:sveihr}(c). The results indicate that the availability of beds and ventilators affects the number of deaths. In the second simulation, we observe a visible spike in the cumulative number of deaths after all the available beds are occupied. As a result, the number of deaths in the first simulation totaled 5,239, and in the second simulation, 28,507 deaths. These illustrative examples demonstrate the potential consequences of resource shortages during a pandemic. They also emphasize the importance of effective resource allocation strategies to reduce mortality rates. 

\begin{table}[]
    \centering
    \small
    \caption{Estimated weekly parameters for \ac{SVEIHR} model.}
    \label{tab:parameters}
    \begin{tabular}{cc|cc|cc|cc|cc}
    \hline
    \textbf{Parameter}  & \textbf{Value} & \textbf{Parameter}          & \textbf{Value} & \textbf{Parameter}       & \textbf{Value} & \textbf{Parameter}            & \textbf{Value} & \textbf{Parameter}            & \textbf{Value}   \\
    \hline
    $\bar{T}$  & 20.00    & $\gamma^v$         & 1.00     & $\mu^\text{ks}$ & 0.70   & $p^\text{c}$         & 0.05 \cite{wu_2020} & $\varsigma^\text{c}$ & 0.4  \cite{wu_2020}   \\
    $\beta$    & 3.20   & $\gamma^\text{m}$  & 1.00    & $\mu^\text{c}$  & 0.35 \cite{zhou_2020}& $p^{\text{r},v}$     & 0.85  & $N$                & 1.2M \\
    $\epsilon$ & 0.95   & $\gamma^\text{s}$  & 0.64 \cite{zhou_2020} & $\sigma$        & 0.64 \cite{zhou_2020}  & $p^{\text{m},v}$     & 0.10   & $E_0$                & 100     \\
    $\rho$     & 0.10   & $\gamma^\text{ks}$ & 0.23 & $p^\text{m}$    & 0.80 \cite{wu_2020}   & $p^{\text{s},v}$     & 0.05  & $I^{\text{m}}_0$     & 50      \\
    $\alpha$   & 0.71 \cite{linton_2020}    & $\gamma^\text{c}$  & 0.33 \cite{zhou_2020} & $p^\text{s}$    & 0.15 \cite{wu_2020} & $\varsigma^\text{ks}$ & 0.40   & $I^{\text{s}}_0$     & 20     \\
    \hline
    \end{tabular}
\end{table}

Recovery ($\gamma$) and mortality ($\mu$) rates vary depending on the severity of the disease and the availability of healthcare resources. The model assumes that critically ill patients transition to the compartment ($\pmb{D}$) when ventilators are unavailable, emphasizing the dire consequences of resource limitations during a pandemic.

\subsection{Modeling of Uncertainties in Vaccine Hesitancy}
\label{sec:uncertainty-paths} 

 Figure~\ref{fig:vh}(a) presents the county-level \ac{VH} estimates derived from the \ac{CTIS} conducted by the Delphi Group at Carnegie Mellon University, in partnership with Facebook \cite{salomone_2021}. Each line in the graph represents the progression of \ac{VH} over time in every county of Arkansas and New York. The graph reveals an overall decline in \ac{VH} rates over time. The graph also reveals various vaccination attitudes across counties in these states. These variations highlight the challenges of achieving widespread acceptance of vaccines across the U.S. For instance, counties in Arkansas exhibit higher \ac{VH} than NY and, consequently, lower vaccination rates  (see Figure~\ref{fig:vh}(b)). This variation is critical because it affects both the spread of the disease and the need for healthcare resources.

\begin{figure}[htb]
    \centering
    \includegraphics[width=0.8\textwidth]{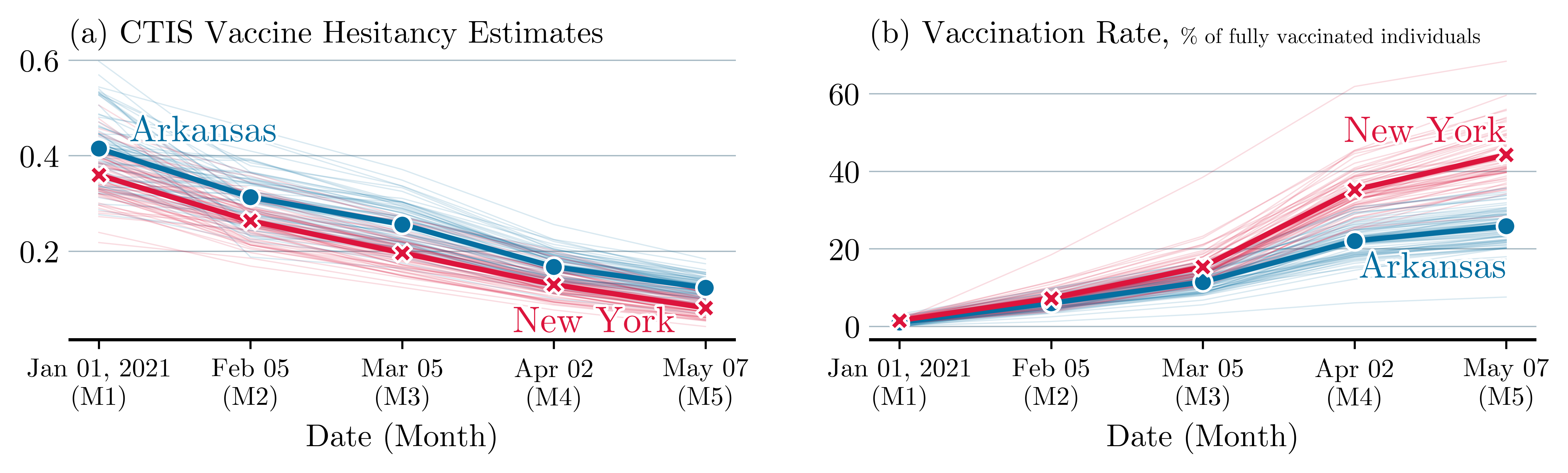}
    \caption{Vaccine hesitancy and vaccination uptake trends of Arkansas and New York counties.}\label{fig:vh}
\end{figure}

We use historical VH estimates from \ac{CTIS} data to determine the rate of change of \ac{VH} from one period to the next. For example, a rate change of -0.15 in period $t$ indicates a 15\% decrease in \ac{VH} from period $t-1$ to $t$. We fit a distribution to the rate of change. Numerical analysis shows that the monthly rate of change of \ac{VH} is normally distributed. The corresponding mean ($\mu_t$) and standard deviation ($\sigma_t$) also change from one month to the next (see Figure ~\ref{fig:roc}(a)-(d)).

To represent the uncertainties associated with \ac{VH}, we develop several scenario paths, represented by $\mathit{VH}^{\omega} \in \Omega$. Each scenario path consists of a series of \ac{VH} values, one per time period,  $\mathit{VH}^{\omega}=\{\mathit{VH}_1^{\omega},\mathit{VH}_2^{\omega}\ldots,\mathit{VH}_T^{\omega}\}$, where $T$ represents the length of the period of study.  We create each scenario path via the following procedure.  Initially, we approximate the normal distribution functions by a triangular, discrete distribution function that takes values $\mu_t, \mu_t - \sigma_t, \mu_t + \sigma_t$ with probability 0.158, 0.684 and 0.158 correspondingly. Next, we create the scenario paths. Each scenario path begins with an initial $\mathit{VH}_1$, which changes by $\mu_t, \mu_t - \sigma_t$, or $\mu_t + \sigma_t$ each period, with probability 0.158, 0.684 and 0.158, correspondingly.  

\begin{figure}[htb]
    \centering
    \includegraphics[width=\textwidth]{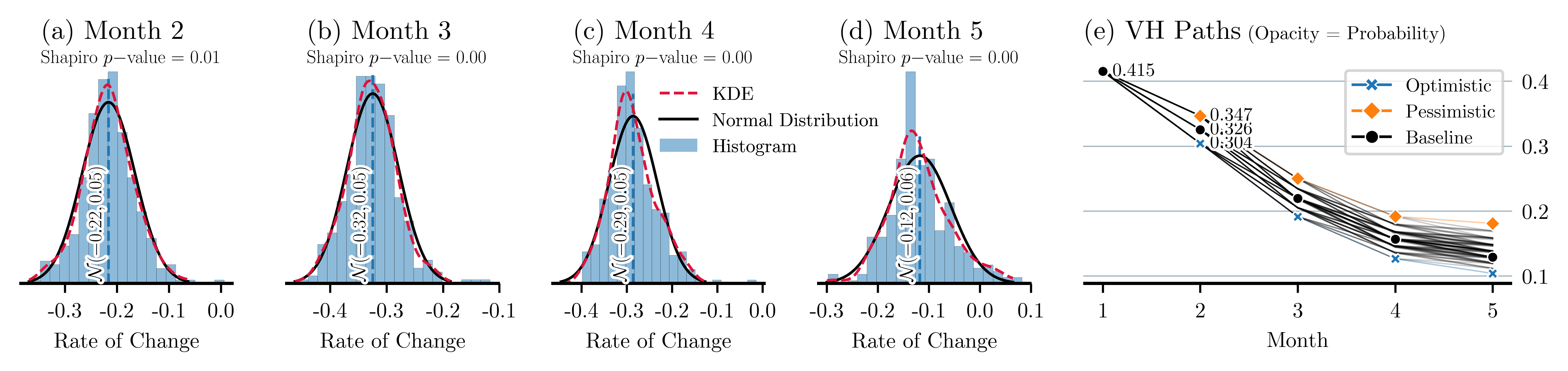}
    \caption{(a)-(d) Distribution of the monthly rate of change of vaccine hesitancy, (e) Example of VH scenario paths over four months.}\label{fig:roc}
\end{figure}

Let us illustrate this via an example. Assume that $\mathit{VH}_1 = 0.415$ and the length of our study period is $T=5$. Using our proposed approach, we construct a total of $3^4=81$ unique scenario paths. Figure~\ref{fig:roc}(e) presents the different scenario paths for this example. Let $p^{\omega}$ represent the probability associated with each scenario path. The graph highlights a baseline, an optimistic, and a pessimistic scenario paths. The optimistic scenario path assumes that the rate of change of VH in each time period is $\mu_t - \sigma_t.$ The pessimistic scenario path assumes that the rate of change of VH in each time period is $\mu_t + \sigma_t.$ As a result, the probability associated with each of these two scenarios is $0.158^4=6.23\times 10^{-4}$}. In the baseline scenario path, the rate of change in each period is $\mu_t,$ and the probability of this scenario path is $0.684^4=2.19\times 10^{-1}$}. 

\section{Numerical Analysis and Results}
\label{sec:results}

\begin{figure}[htb]
    \centering
    \includegraphics[width=0.8\textwidth]{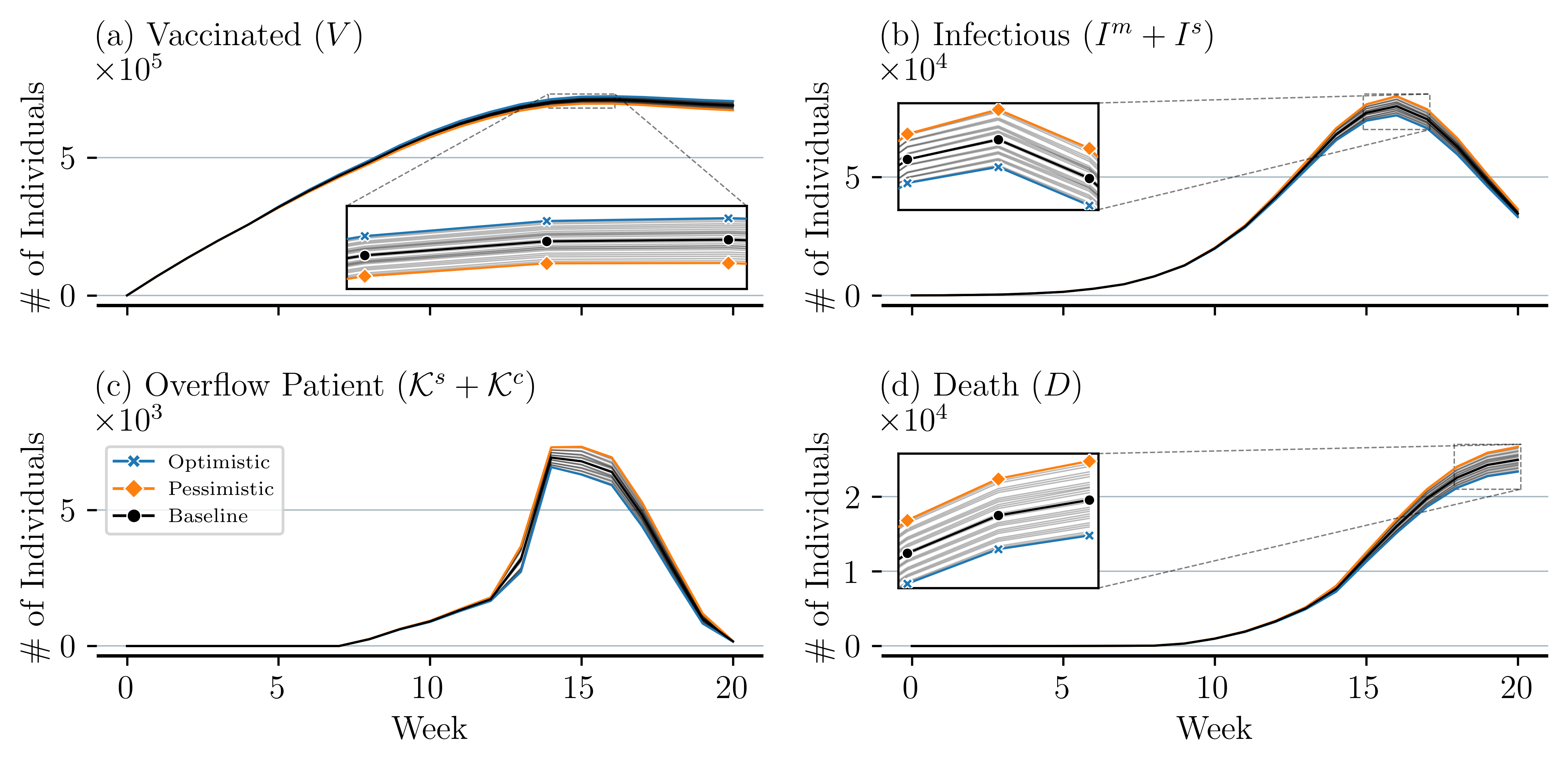}
    \caption{Evaluating the impact of different scenario paths illustrated in Figure~\ref{fig:roc}(e) on the outcomes of SVEIHR model. The optimistic, pessimistic, and baseline scenario paths are highlighted.}\label{fig:exp1}
\end{figure}

Our numerical analysis uses historical VH estimates from \ac{CTIS} data for five months, from January 1 to May 7, 2021. We focus our study on this period because COVID-19 vaccines were available in the U.S. Via our numerical analysis, we address three important research questions. To address {\bf RQ1:} \emph{How do changes in public attitudes towards vaccination affect epidemic outcomes?} We simulate the \ac{SVEIHR} model weekly for 20 weeks (5 months). We conduct a total of $|\Omega|=81$ simulations, one for each $\mathit{VH}^{\omega} \in \Omega$, and summarize the results in Figure~\ref{fig:exp1}. 

In Figures~\ref{fig:exp1}(a) to ~\ref{fig:exp1}(d), the blue line represents the optimistic VH scenario path, which assumes a continuous decrease in VH, resulting in the highest vaccination uptake. The orange line represents the pessimistic VH scenario path, for which vaccination uptake is lowest. The results of Figure~\ref{fig:exp1}(b) indicate that high vaccination rates lead to a noticeable decrease in the number of infections. Results in Figure~\ref{fig:exp1}(c) indicate that high vaccination rates reduce the number of patients whose needs for healthcare resources (such as beds and ventilators) are not met. The death toll for the scenario paths with higher vaccination rates is lower compared to those with lower vaccination rates (see Figure~\ref{fig:exp1}(d)).  The maximum difference in vaccination among the different scenario paths, at the end of the study period, is 33,099 cases. The corresponding maximum difference in the number of infected is 8,079, the maximum difference in the number of patients whose needs for healthcare resources were not met is 1,023, and the maximum difference in the number of deaths is 3,323. These results highlight the significant influence of \ac{VH} on public health outcomes during a pandemic. Reducing \ac{VH} is linked to better health outcomes, which emphasizes the critical role of \ac{VH} mitigation in controlling the spread of infectious diseases.

\begin{figure}[htb]
    \centering
    \includegraphics[width=\textwidth]{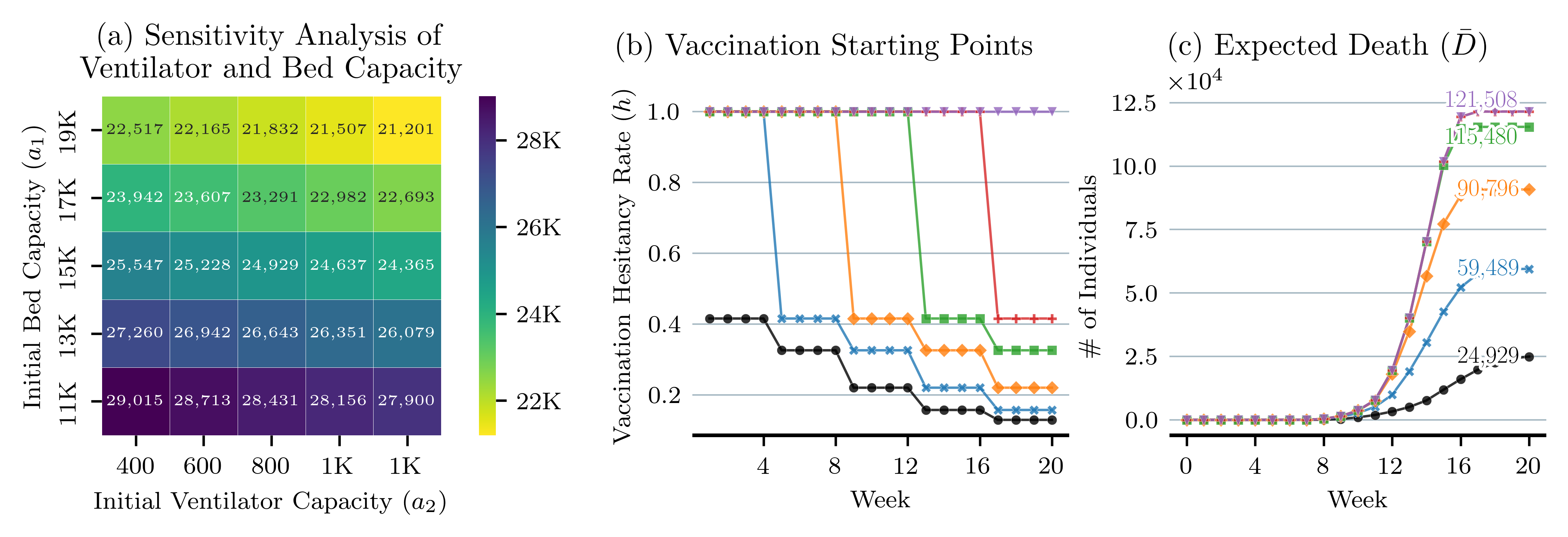}
    \caption{(a) Expected number of fatalities relative to the number of beds and ventilators available. (b) Changes in weekly VH over time for the baseline scenario path. (c) Expected number of fatalities over time.}\label{fig:exp2}
\end{figure}

To address {\bf RQ2:} \emph{How do healthcare resources impact patient outcomes during a pandemic?} We design a total of 25 problems by varying the number of ventilators and beds available. For each of these problems, we run a total of 81 simulations, each corresponding to a different scenario path $\mathit{VH}^{\omega} \in \Omega$. Each simulation uses the \ac{SVEIHR} model and runs for 20 weeks (5 months). We summarize the results in Figure~\ref{fig:exp2}(a). The numbers represent the expected number of fatalities for each problem. The results reflect an increase in fatalities as the availability of resources decreases, highlighting the direct correlation between healthcare infrastructure and survival rates. The results show that maintaining a robust supply of beds and ventilators is essential for keeping fatalities to a minimum.

We develop a regression model using the data from the experiments. The independent variables for this model are the number of beds ($a_1$) and ventilators ($a_2$), and the dependent variable is the expected number of fatalities ($\bar{D}$). Based on model~\eqref{eq:linear}, for every available bed ($a_1$), the expected number of deaths reduces by 0.827, and every additional available ventilator ($a_2$) reduces the expected number of deaths by 1.511.

\begin{equation}
    \bar{D} = 38657.33 - 0.827a_1 - 1.511a_2 \label{eq:linear}
\end{equation}

To address {\bf RQ3:} \emph{How do vaccine rollout times affect the progression and severity of the epidemic?} We create a total of 6 problems by changing the timing of vaccine rollouts. One of the problems assumes that vaccines were available at the beginning of the study period. Each of the remaining problems assumes a delay of 1 or 2, \ldots, or 5 months in vaccine rollout. For each problem, we run a total of $|\Omega|$ simulations, one for each scenario path $\mathit{VH}^{\omega}$. Each simulation runs the \ac{SVEIHR} model for 20 weeks (5 months). We summarize the results in Figures~\ref{fig:exp2}(b) and (c).  

In our model, we control vaccination rollout via \ac{VH} rate, $h$. Figure~\ref{fig:exp2}(b) presents the changes in $h$ over time for each problem, for the baseline scenario path only. The changes in $h$ for the baseline scenario path are presented by the black line. If no one is vaccinated during the study period, VH is 100\%. This problem is represented by the straight line at $h=1.$ If vaccine rollout is delayed by 1 month, $h=1$ for the 1st month, and next, $h$ decreases continuously as people are vaccinated. The specific decrease in $h$ over time depends on the specific scenario path. In Figure~\ref{fig:exp2}(b), the changes in $h$ follow the changes of the baseline scenario path.

Figure~\ref{fig:exp2}(c) presents the expected number of fatalities for each problem. The results show that the earliest vaccinations begin, the highest the decrease in fatalities. The expected number of fatalities at the end of the study period is 4.88 times higher if vaccines were not available versus vaccines being available at the start of the study period.  We notice that the difference in the expected number of fatalities when vaccines are rolled out in the last month versus vaccines never being available is statistically insignificant. These results underscore the life-saving potential of prompt and efficient immunization programs.   

\section{Conclusions and Future Research Directions}
\label{sec:conclusion}

{\small This study presents an extension of compartmental epidemiological models developed to estimate the spread of the COVID-19 virus and its impacts on the population. The proposed model captures the influence of vaccine availability, \ac{VH}, and the availability of healthcare resources on disease dynamics, the expected number of hospitalizations, and the expected number of fatalities. We simulate the proposed model for different \ac{VH} scenario paths and different problem settings. Our analysis highlights the significant influence of ($i$)  VH on public health outcomes during the pandemic; ($ii$) critical healthcare resources on the expected number of fatalities; and ($iii$) timing of vaccine rollouts on the expected number of fatalities. 

Our data analysis shows that \ac{VH}  changes over time and regions of the U.S. The rate of change in \ac{VH} from one period of time to the next is normally distributed. The proposed simulations of \ac{SVEIHR} model highlight the impact of stochastic \ac{VH} on model outcomes. Our next steps will focus on extending this work by developing a \acf{MSP}  that supports resource allocation decisions under uncertainty. The proposed model will consider budget limitations and will ensure fairness within the healthcare system. We plan to use an \ac{MSP} because it facilitates sequential decision-making over time and effectively addresses resource availability and evolving uncertainties of \ac{VH}, infection rate, etc. Such models mimic reality since decision-makers continually refine their decisions as new data emerge to ensure a proactive and adaptive response to a pandemic.   

\vspace{-0.2in}
\bibliography{references.bib}

\begin{acronym}
    \acro{OR}{operations research}
    \acro{SIR}{susceptible-infectious-recovered}
    \acro{SEIR}{susceptible-exposed-infectious-recovered}
    \acro{SVEIHR}{susceptible-vaccinated-exposed-infectious-hospitalized-recovered}
    \acro{MSP}{multi-stage stochastic program}
    \acro{VH}{vaccine hesitancy}
    \acro{CTIS}{COVID-19 Trends and Impact Survey}
    \acro{ODE}{ordinary differential equation}
\end{acronym}

\end{document}